\documentclass[conference]{./sty/IEEEtran}
\usepackage{sty/packages}

\begin{document}
\newcommand{\toniot}{TON\_{}IoT}
\newcommand{\TP}{\mathit{TP}}
\newcommand{\TN}{\mathit{TN}}
\newcommand{\FP}{\mathit{FP}}
\newcommand{\FN}{\mathit{FN}}
%
\title{Intrusion Detection in Internet of Things\\ using Convolutional Neural Networks}
%
%
\author{\IEEEauthorblockN{Martin Kodyš, Zhi Lu, Kar Wai Fok, and Vrizlynn L. L. Thing}
\IEEEauthorblockA{Cyber Security Strategic Technology Centre, ST Engineering\\
Singapore}}


%


\maketitle


\begin{abstract}

Internet of Things (IoT) has become a popular paradigm to fulfil needs of the industry such as asset tracking, resource monitoring and automation.
As security mechanisms are often neglected during the deployment of IoT devices, they are more easily attacked by complicated and large volume intrusion attacks using advanced techniques.
Artificial Intelligence (AI) has been used by the cyber security community in the past decade to automatically identify such attacks.
However, deep learning methods have yet to be extensively explored for Intrusion Detection Systems (IDS) specifically for IoT.
Most recent works are based on time sequential models like LSTM and there is short of research in CNNs as they are not naturally suited for this problem.
In this article, we propose a novel solution to the intrusion attacks against IoT devices using CNNs.
The data is encoded as the convolutional operations to capture the patterns from the sensors data along time that are useful for attacks detection by CNNs.
The proposed method is integrated with two classical CNNs: ResNet and EfficientNet, where the detection performance is evaluated.
The experimental results show significant improvement in both true positive rate and false positive rate compared to the baseline using LSTM.

\end{abstract}


%
\IEEEpeerreviewmaketitle

\section{Introduction}

Internet of Things (IoT) interconnects billions of devices and keeps growing\cite{ahmad2021machine}.
Within the IoT, all kinds of objects communicate without human intervention\cite{alaba2017iot-surv}.
By reading data from sensors, or controlling actuators remotely, IoT enables services ranging from home automation to optimisation of complex workflows in the industry. Industrial Internet of Things (IIoT) 
connects sensors and controllers similarly to the established Supervisory Control And Data Acquisition (SCADA) systems\cite{daneels1999scada}, but provides wider connectivity and interoperability\cite{samtani2018identifying}.

Despite the great potential of IoT and IIoT, economically successful devices often lack a robust security implementation. This is especially alarming in industrial applications. A compromised IIoT system may lead to economic loss, physical damage to equipment or goods, and can even cause bodily harm to people.

Along protective measures, cyber security involves the recognition of potential threats. An Intrusion Detection System (IDS) is a software component that monitors and analyses various activities and measures of the system to detect attacks.

The complexity and quantity of attacks push for more efficient IDSs. Although the traditional machine learning provides quick processing, its design is slowed down by the manual feature engineering for each new threat. In contrast, the deep learning brings an end-to-end approach combining feature selection and classification\cite{rezaei2019deep}. Its automation speeds up the defence response against the fast-evolving cyberattacks.

Numerous approaches implement deep learning on well-known network datasets. 
A Convolutional Neural Network (CNN) GoogLeNet with Deep Random Forest was presented in \cite{zhang2019multiple}. The IDS proposed in \cite{kim2020ai} based on CNN-LSTM architecture detects variants of attacks unrecognised by signature-based IDSs, effectively reducing the false alarm rate. Another IDS, proposed in \cite{shone2018dl-ids}, uses stacked non-symmetric deep Auto-Encoders coupled with Random Forest.

Lacking IoT/IIoT data\cite{al-hadhrami2020dataset-framework} in current datasets motivated the introduction of \toniot{} Telemetry dataset in \cite{alsaedi2020ton_iot}. Its authors present an intrusion detection based on IoT/IIoT sensor readings, and provide a baseline evaluation of traditional machine learning techniques including Random Forests, Naïve Bayes, and Support Vector Machines, and a deep learning technique using Long Short-Term Memory (LSTM).


Our contribution in intrusion detection is a novel use of classical CNNs on sensor readings with a time-based encoding leveraging on the visual patterns of missing values.

Akin to slit-scan photography, stacked sensor data represent an evolution of a state in time. Our method thus enables a CNN to directly capture temporal patterns.

The paper is structured as follows. \Autoref{sec:related_works} provides a review of related works and recent progress in the area of attack detection in IoT. \Autoref{sec:methodology} presents the different steps of our approach and details its implementation. \Autoref{sec:experiment_design} explains the experiment design. In \Autoref{sec:results}, results of the experiments are presented. \Autoref{sec:discussion} discusses the pros and cons of our proposed approach and suggests future research directions.



\section{Related Works} 
\label{sec:related_works}
This section presents cyber security challenges in IoT/IIoT, related attack types, and existing works in IoT/IIoT security using machine learning and deep learning.

\subsection{Internet of Things}
\label{sub:iot_iiot}
	Over the past 5 years, the interest in securing IoT has seen an increasing trend indicated by a growing number of intrusion detection surveys that either include IoT\cite{berman2019,gumusbas2021} or also focus specifically on IoT security \cite{alaba2017,zarpelao2017,al-garadi2020,khraisat2021}.
	The increased vulnerability of IoT devices stem from their limited resources (memory, power supply, processing power, bandwidth).
	The survey \cite{berman2019} highlights challenges in the detection of botnets, compromised devices attacking other targets.
	Others include real-time attack detection \cite{ahmad2021machine}; improvement of validation methods \cite{zarpelao2017}; availability of training datasets and their efficient use, zero-day attacks, integration with other technologies (e.g. blockchain) \cite{al-garadi2020}; and dealing with evasion techniques\cite{khraisat2021}.

\subsection{Attacks and Detection} 
\label{sub:attacks_and_detection}

Common IoT attacks, such as password attacks via Recurrent Generative Adversary Networks (GANs)\cite{nam2020password}, expose the lack of secure device configuration and use of default passwords.
Jamming and Denial of service (DoS) attacks prevent correct functioning of devices or disrupt the stream of information critical for some processes. Jamming was addressed in\cite{namvar2016jamming} with a game-theory perspective using evolutionary algorithm to maximise chances of getting IoT signal through. DoS attack floods the victim with requests and can slow down, induce crash, or shutdown the target. \cite{galeano-brajones2020dos} proposes an entropy-based solution in Software Defined Networking (SDN).

In a Man-in-the-Middle attack, the malicious agent covertly intercepts and alters the communication. \cite{wong2020mitm} introduces an attack on MQTT exchange using an adversarial message generation.

Simple web servers are often a user interface for IoT devices, or the dashboard for their management. Therefore, cross-site scripting (XSS)\cite{owasp_xss} on web resources that do not properly sanitise their inputs, triggering of a broken session management, or having deficiencies in their application logic\cite{ayeni2018xss} are of concern.

Scanning (or reconnaissance) attack aims at detecting the characteristics of the network and devices in order to exploit their weaknesses. For instance, \cite{al-alami2017scanning} makes use of Shodan\cite{shodan}, an online database, to find potentially vulnerable devices.

Zero-day attacks are characterised by their novelty and can be of any of previously mentioned types. They present a serious threat as the attack vector had not been described before their deployment.
To deal with them, \cite{sharma2018framework} proposes a context-graph model to support zero-day attack detection, response and re-establishment of trust.
A detection using DL was also described in\cite{hindy2020utilising} which uses auto-encoders on CICIDS2017 network capture dataset.

From another perspective, an anomalous power consumption of IoT devices can reveal an attack on the network \cite{myridakis2020iot-power-id}.

\subsection{Machine Learning for IoT/IIoT Security} 
\label{sub:machine_learning_in_iot_iiot_protection}


Deep Learning (DL) is becoming an unavoidable part of the attack detection toolset.
Compared to shallow neural networks, deep neural networks are able to detect higher-level concepts. Competitiveness of deep learning improved with an increased Graphical Processing Units (GPUs) availability and parallel computing of layers that make the complex computations faster\cite{berman2019}.

On IIoT, \cite{zolanvari2019ml-iiot} offers a case study and evaluates several ML techniques on SCADA system in a testbed setting.
Although \cite{abu--al-haija2020} proposed a DL approach using CNNs intended for detection of malicious traffic in IoT networks, the experiments were conducted on the NSL-KDD dataset\cite{NSL-KDD} which contains general network traffic, not specific to IoT. Their approach reshapes the features of a single record into a square matrix that is then used as the input to a CNN, while in our approach the features of a single record form only one row of the input matrix.

Aimed at IoT botnet detection a system in \cite{bezerra2019iotds} performs a continuous analysis based on several ML methods, which include Elliptic Envelope, Isolation Forest, Local Outlier Factor, and One-class Support Vector Machine. They evaluated the impact of their solution in terms of memory and CPU load.

An example of a successful use of CNNs in cybersecurity is malware detection described in \cite{wang2017cnn-malware}. This work uses network traffic merged from various public datasets. The input pictures are obtained from network captures of data flows. Such a picture is a single-channel bitmap snapshots trimmed to first $784$ bytes. As the flow indicates only the order and no temporal features, the authors suggest that their future work will seek to represent the data as time series. Our approach adapts time-series directly in the encoding as a dimension in the input tensor for the CNNs.

CNNs were incorporated in network traffic intrusion detection as a part of an ensemble method applied to Ethernet Consist Network \cite{Yue2021ensemble-trainECN}. However, the temporal aspect of the data was addressed by another component based on LSTM. Similarly, \cite{Xiao2019CNN-IDS} uses CNNs uniquely on a set of features reshaped in a two-dimensional array. In our approach, CNNs are applied on "slit-scan" images to capture the temporal links.




In \cite{alsaedi2020ton_iot}, the authors formulated the need of an IoT/IIoT dataset that was very much lacking. The proposed dataset is \toniot{}, where readings of IoT sensors are collected and provided. It is in response to the missing sensor data among currently available datasets. In \cite{alsaedi2020ton_iot}, the authors also provided a baseline evaluation of supervised ML methods, which include logistic regression, linear discriminant analysis, $k$-Nearest Neighbours, Random Forest, Classification and Regression Trees, Naïve Bayes, Support Vector Machine, and LSTM, on the \toniot{} dataset.
In this paper, we will use the \toniot{} dataset for our proposed work.



\section{Methodology} 
\label{sec:methodology}
Our aim is to use sensor readings to detect attacks in an IoT/IIoT system. The process starts from reading the annotated multiple sensor sources (\Autoref{sub:dataset_exploration}) and prepares the data for deep learning (\Autoref{sub:data_processing}). Deep learning models are then used to predict whether there is an attack and if so, determine its type.


\subsection{TON IoT Dataset} 
\label{sub:dataset_exploration}

In 2019, \toniot{} dataset was created based on a testbed environment at the Cyber Range and IoT Labs at the University of New South Wales (UNSW) Canberra, Australia\footnote{UNSW Canberra: \url{https://www.unsw.adfa.edu.au}}. The experimental setting described in \cite{alsaedi2020ton_iot} comprises a combination of physical and simulated connected devices. The dataset name "\toniot" refers to its contents and provenance: Telemetry data, Operating systems’ data, and Network data from
the IoT/IIoT testbed.

The testbed was organised into three layers connected with virtualised switches and routers: Cloud layer (virtualised), Fog layer (virtualised), and Edge layer (physical).

\textit{Cloud layer} contains a Message Queuing Telemetry Transport (MQTT) broker. \textit{Fog layer} contains the monitored vulnerable hosts and attacker machines. It includes a Node-RED \cite{nodered} server that simulates 6 out of 7 sensor sources. \textit{Edge layer} contains the physical devices, from which an ESP8266 weather station acts as the \nth{7} sensor data source. Other devices in the edge layer are two smartphones and a smart TV (although mentioned in \cite{alsaedi2020ton_iot}, they are not contributing any IoT data), and the NSX VMware server emulating virtualised layers. 

\begin{figure}[!t]
  \includegraphics[width=.8\columnwidth]{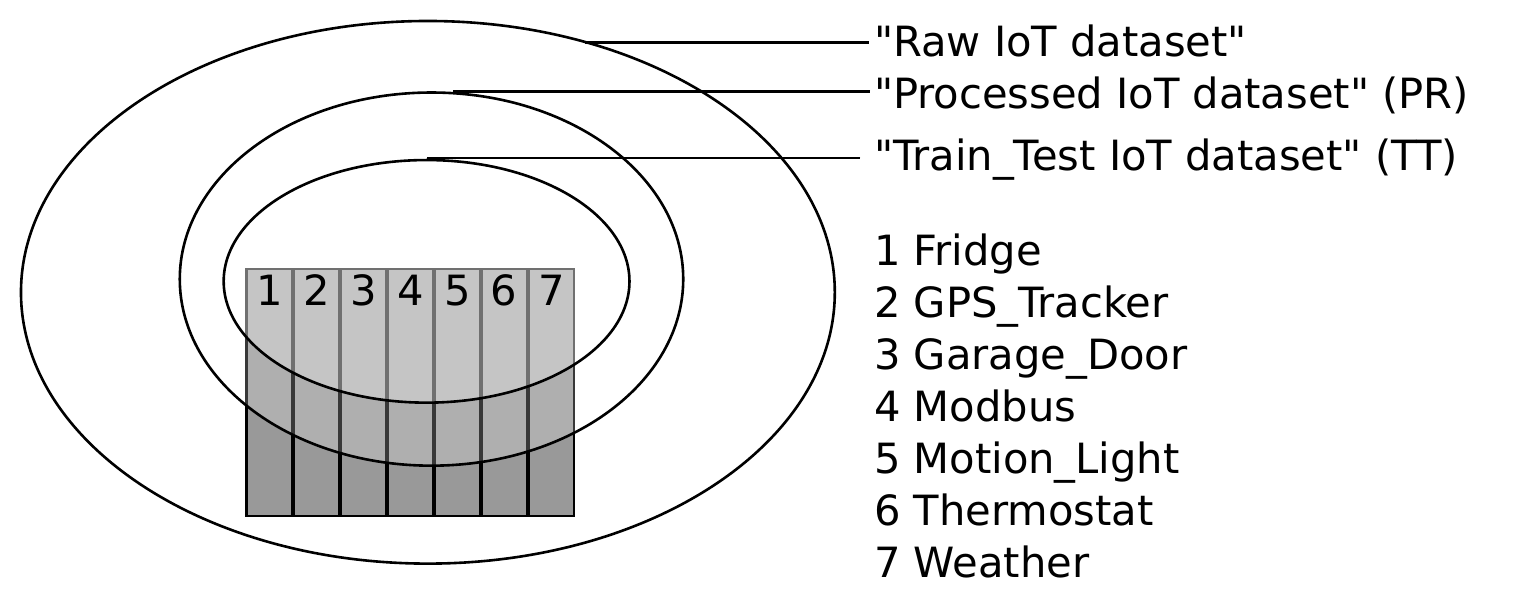}
  \centering
	\caption{IoT telemetry datasets within \toniot{} dataset. Datasets used in our paper are "Train\_Test IoT dataset" referred to as "\textit{TT dataset}", and its superset "Processed IoT dataset" referred to as "\textit{PR dataset}". The numbered items represent the data from 7 sensors.}
	\label{fig:datasets}
\end{figure}

Aiming at capturing the specifics of IoT and IIoT systems, we focus on the IoT telemetry part of this dataset. \toniot{} telemetry dataset provides the sensor readings as registered by a Node-RED server, delivered through the MQTT protocol via the MQTT broker in the Cloud layer. This part of data is provided in the form of comma-separated values (CSV) files, one per sensor. 
The content of the \toniot{} dataset is illustrated in \autoref{fig:datasets}. 
Along with the IoT telemetry data, the dataset also contains network captures and process data from operating systems on targeted virtual machines, which are out of scope of our paper.

\subsubsection{Data Overview} 
\label{ssub:data-overview}

The selected \toniot{} dataset is highly heterogeneous. We provide a closer look at its content for a better understanding of its characteristics.

The data is of these categories: IoT telemetry data (i.e. IoT/IIoT sensor readings, on which we focus in this paper), network data, Linux host data, and Windows host data. Additionally, there is a specification of ground truth for the security events.
The dataset also contains basic statistics about the number of records in each category.

\autoref{fig:datasets} shows a hierarchical structure of the subsets in the IoT telemetry dataset, which are "Raw", "Processed", and "Train\_{}Test", respectively. For clarity, we will refer to them by aliases. \toniot{}'s "Processed IoT dataset" will be referred to as "PR dataset", \toniot{}'s "Train\_Test IoT dataset" will be called "TT dataset". "Train\_Test dataset", despite its name, actually does not contain a dataset split into Train and Test datasets. Instead it is a single dataset that its authors mention to be used for training and testing in $80:20$ proportion. The exact Train and Test datasets used in evaluation in \cite{alsaedi2020ton_iot}, were not provided by the authors.

The \textit{Raw} dataset contains the source data in different formats (e.g. log, JSON) depending on sensor type. Its conversion into CSV format produces the \textit{PR} dataset where each of 7 sensors has the data in a separate CSV file. The \textit{TT} dataset, composed of 7 CSV files as well, is a subset of the records from the \textit{PR} dataset. The \textit{TT} dataset contains an additional "ts" (timestamp) field, that corresponds to "date" and "time" fields converted to UNIX epoch time in Pacific Daylight time zone (UTC-0700).

The telemetry data come from 7 sensors. Six sensors were simulated in Node-RED framework. The seventh sensor, ESP8266 weather station, was physically connected to the testbed infrastructure. 
A summary of the sensors and their output values is given in \autoref{tab:sensor-ranges}.

\begin{table}[!t]
	\renewcommand{\arraystretch}{1.3}
	\caption{Sensor ranges in TT dataset}
	\label{tab:sensor-ranges}
	\centering
	\begin{tabular}{ r l l }
		\hline
		Sensor & Measure & Values \\ \hline
		Fridge &
      fridge\_{}temperature & [1.0; 14.0] \\
      &temp\_{}condition & \{'high', 'low'\}\\
              
		Garage\_{}Door &
      door\_{}state & \{'open', 'closed'\}\\
      &sphone\_{}signal & \{'true', 'false', 0, 1\}\\
              
		GPS\_{}Tracker &
      latitude &[0.00; 550.00]\\
      &longitude & [10.00; 556.00]\\
              
		Modbus &
      FC1\_{}Read\_{}Input\_{}Register & $\mathbb{N} \cap [0; 65535]$ \\
			&FC2\_{}Read\_{}Discrete\_{}Value & $\mathbb{N} \cap [0; 65535]$ \\
      &FC3\_{}Read\_{}Holding\_{}Register & $\mathbb{N} \cap [0; 65535]$ \\
      &FC4\_{}Read\_{}Coil & $\mathbb{N} \cap [0; 65535]$ \\

		Motion\_{}Light &
      motion\_{}status & \{0, 1\}\\
			&light\_{}status & \{'on', 'off'\}\\
              
		Thermostat &
      current\_{}temperature & [25.00; 35.00] \\
  		&thermostat\_{}status & \{0, 1\} \\
		
		Weather &
    	temperature & [20.00; 50.00] \\
      &pressure & [-12.00; 12.50] \\
      &humidity & [0.20; 100.00] \\
\hline
	\end{tabular}
\end{table}

Both datasets (\textit{PR} and \textit{TT}) offer 17\,features from 7\,sensors along with a date and time indication. Every data recording was annotated with a binary \texttt{label} field (\texttt{0} means normal operation, \texttt{1} indicates an attack). In addition, the \texttt{type} field specifies the event type: if label = 0, \texttt{normal}, otherwise it indicates \texttt{backdoor},  \texttt{ddos}, \texttt{injection}, \texttt{password}, \texttt{ransomware}, \texttt{scanning}, or \texttt{xss}.


\subsubsection{Attack distribution} 
\label{ssub:attack_distribution}

\begin{figure}[!htbp]
  \includegraphics[width=\columnwidth]{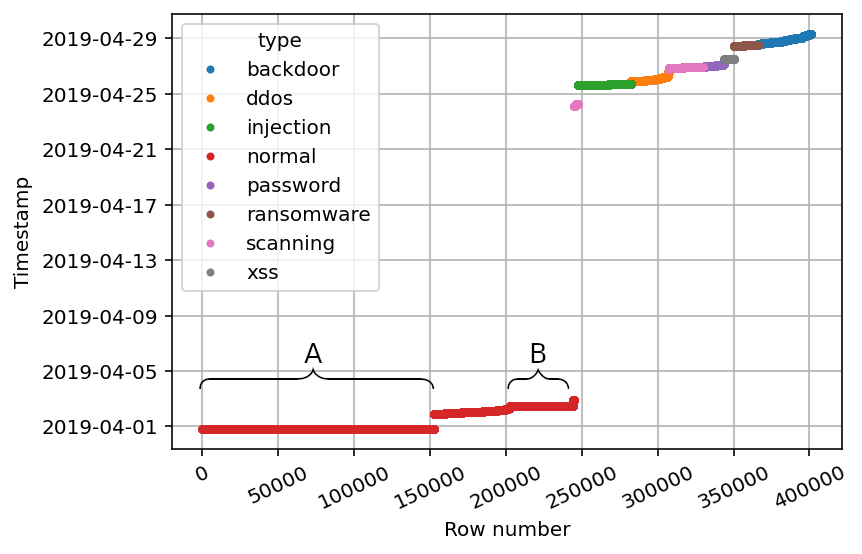}
  \centering
  \caption{Composition of concatenated \textit{TT} dataset conform to description in \cite{alsaedi2020ton_iot} of the "Combined\_IoT\_dataset", "A" and "B" mark segments where many readings have the same timestamps.}
  \label{fig:composition-tt}
\end{figure}

\autoref{fig:composition-tt} shows a major data interruption that separates the TT dataset in two. The first part of the dataset, collected between \nth{31} March and \nth{2} April, has all data points marked as normal. The second part of the dataset, from \nth{24} to \nth{30} April, was marked as different attacks. To prevent exploitation of this characteristics of the \toniot{} datasets by our method, the fields related to time and date were excluded from the training process.

\subsubsection{Multiple Readings Issue} 
\label{ssub:multiread-issue}
\begin{figure}[!htbp]
  \includegraphics[width=\columnwidth]{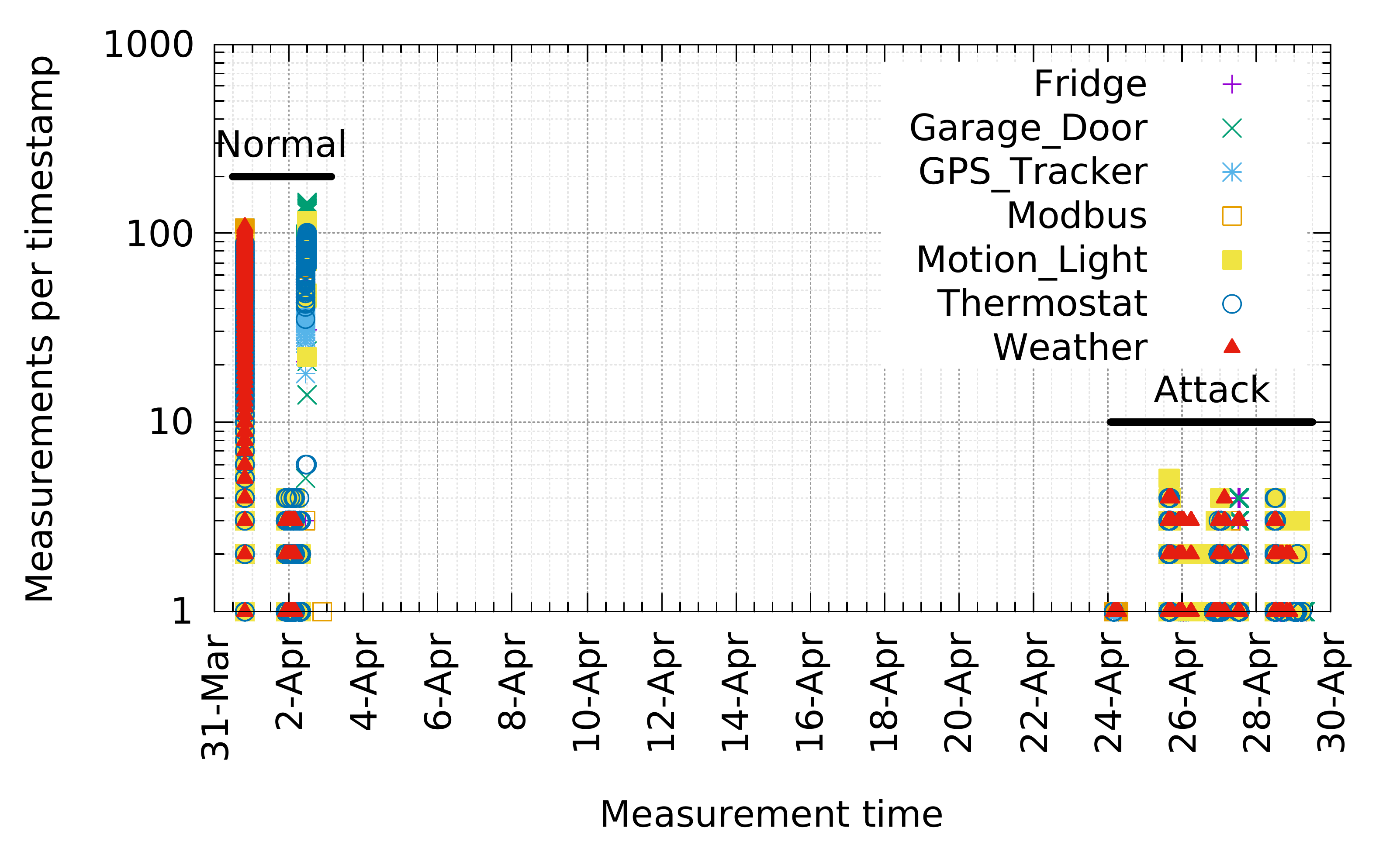}
  \centering
  \caption{Number of sensor events per timestamp for each sensor in the TT dataset. The dataset timestamps have granularity of 1\,second. Many \texttt{normal} class events are marked with the same timestamp.}
  \label{fig:counts}
\end{figure}

Visual representation of the data allows us to capture anomalies of the particular set of data and prepare the processing pipeline accordingly. The TT dataset (as well as its superset PR) is characteristic of multiple readings with the same timestamp for the same sensor, as illustrated in \autoref{fig:counts}. We observe peaks of readings per second for all sensors: 145 readings per second from Garage Door sensor, 117 from Motion Light sensor, 115 from Fridge sensor, 108 from Weather sensor, 106 from both Modbus and GPS Tracker sensors, and 100 from Thermostat sensor. All of these readings are labelled as \texttt{normal}. This particularity of the dataset was not explained in the \toniot{} dataset's documentation or its corresponding paper\cite{alsaedi2020ton_iot}. As proposed in \Autoref{ssub:aggregation}, a time-based aggregation can be used to filter a potentially overwhelming quantity of data.

In the \textit{TT} dataset, there is a total of \num{401119} readings from all sensors, which were recorded having \num{85664} unique timestamps. In an ideal data processing case, all sensors would provide a reading at the same time, and timestamps would be unique within each sensor's readings. In reality, neither condition is satisfied. The original distribution of different classes of attack is shown in \autoref{fig:composition-tt}. After removing duplicate entries with respect to date-time, i.e. a single data point represents a given moment, segments marked as "A" and "B" would reduce to a few points. The most visible contraction of the data is that the first \num{150000} records, marked in the figure as "A" (almost a third of the TT dataset) spread only across less than \num{2000} different timestamps representing less than 8\,minutes.

The removal of redundant timestamps affects the balance of different classes in the dataset. As the most of the duplicate timestamps are in the \texttt{normal} class, its proportion drops from 61\,\% to only 25\,\%. The \texttt{backdoor} class becomes the dominant class with almost 26\,\%, as shown in the summary \autoref{tab:aggregation-counts}. As the models tend to be biased towards the majority class, a bias towards the \texttt{backdoor} class is to be expected when using the datasets with this distribution.

\begin{table}[!t]
	\renewcommand{\arraystretch}{1.3}
	\caption{Effects of removal of timestamp duplicity on the TT dataset}
	\label{tab:aggregation-counts}
	\centering
	\begin{tabular}{ r r r r r }
		\hline
Class  &  \multicolumn{2}{c}{Original TT dataset} &	\multicolumn{2}{c}{Unique timestamps in TT} \\ \hline

backdoor &	\num{35000} &	(8.73\%)  & 		\num{25042} &	(29.23\%) \\
ddos &	\num{25000} &	(6.23\%)  & 		\num{11520} &	(13.45\%) \\
injection &	\num{35000} &	(8.73\%)  & 		\num{6524} &	(7.62\%) \\
\textbf{normal} &	\num{245000} &	(61.08\%)  & 		\num{22179} &	(25.89\%) \\
password &	\num{35000} &	(8.73\%)  & 		\num{12924} &	(15.09\%) \\
ransomware &	\num{16030} &	(4.0\%)  & 		\num{4170} &	(4.87\%) \\
scanning &	\num{3973} &	(0.99\%)  & 		\num{2016} &	(2.35\%) \\
xss &	\num{6116} &	(1.52\%)  & 		\num{1289} &	(1.50\%) \\
\hline
Total & \num{401119} & & \num{85664} \\

	\end{tabular}
\end{table}


\subsection{Data processing} 
\label{sub:data_processing}

\begin{figure}[!htbp]
  \includegraphics[width=\columnwidth]{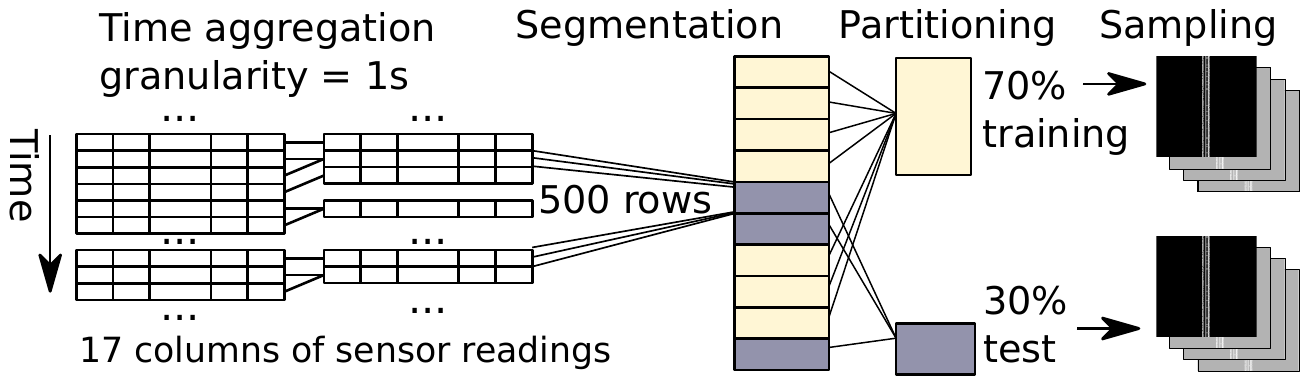}
  \centering
  \caption{Steps of the process include aggregation that filters only one event per second (the maximum resolution of the data), segmentation to keep consecutive readings together, partitioning of the dataset to non-overlapping training and testing subsets, sampling encodes the data into a tensor serving as the input to CNNs.}
  \label{fig:process}
\end{figure}

The data processing method proposed in this paper converts the time-based one-dimensional data from the \textit{PR} dataset (alternatively from the \textit{TT} dataset) into tensors that are accepted by CNNs. The workflow of this method is illustrated in \autoref{fig:process}, which consists of the following steps: \textit{combination}, \textit{aggregation}, \textit{segmentation}, \textit{partitioning}, \textit{imputation}, \textit{tensor construction}, before serving the data to CNNs for training and testing.

Specifically, the \textit{combination} puts various sensor readings into a single dataset. The \textit{aggregation} deals with the issue of multiple readings from a sensor recorded with the same timestamp. The \textit{segmentation} divides the whole dataset into a set of chunks to preserve temporal contiguity of sensor readings. These chunks are then randomly dispatched into either training set or test set by the \textit{partitioning} operation. They are used to compare the performance of various proposed strategies. An \textit{imputation} strategy defines how missing values, that appear during the combination step, are interpreted. The \textit{tensor construction} selects sequences of records, applying the imputation strategy on them to form a three-channel image-like tensor as the input to the CNNs.

\subsubsection{Combination} 
\label{ssub:combination}
The combination step takes seven sensors' source CSV files as inputs and outputs a fusion of the records in a single object that contains all features in columns.

\begin{figure}[!htbp]
  \includegraphics[width=.9\columnwidth]{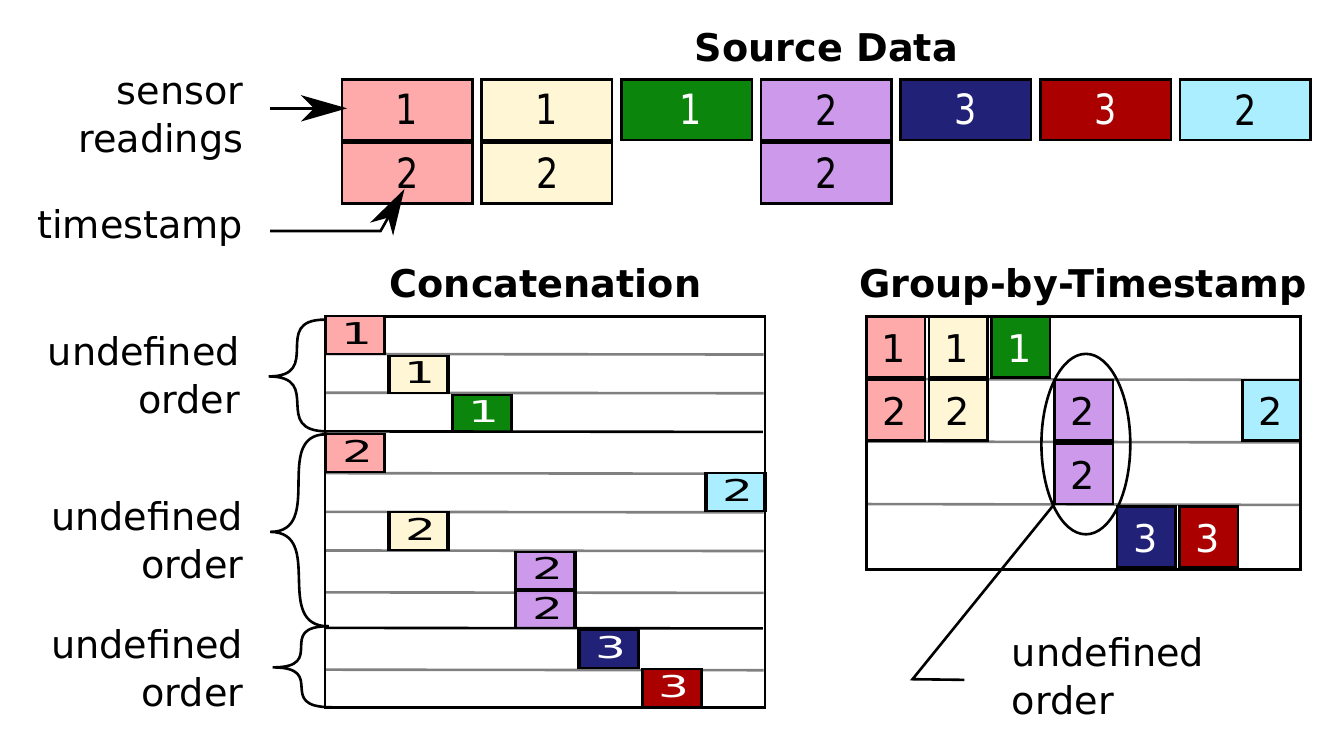}
  \centering
  \caption{Different ways to combine the data sources. Group-by-timestamp provides more dense and less ambiguous representation. Concatenation provides sparse representation with more variability in the ordering of same-timestamp records. Note that one sensor reading can contain more than one variable.}
  \label{fig:combination}
\end{figure}

Two approaches to combining the sensor data were explored are illustrated in \autoref{fig:combination}: \textit{concatenation} and \textit{group-by-timestamp}. \textit{Concatenation}, implemented in \cite{alsaedi2020ton_iot}, places different sensor readings on separate rows, leaving other sensor's values unassigned. \textit{Group-by-timestamp} approach combines sensor readings having the same timestamp within one row. In case of multiple rows having the same timestamp within one source (i.e. sensor), each row has an \texttt{order} number assigned, which can be used to join several readings without aggregating the result. As a result, the dataset can contain multiple readings for one timestamp. The representation is more dense (less unassigned measurement values) and takes into account the temporal co-occurrence of different values as they are combined into one data point. An issue with redundant timestamps is that there is no other ordering indication other than order of appearance in the original dataset (CSV file).

\subsubsection{Aggregation} 
\label{ssub:aggregation}
The aggregation takes a table-like object and contracts rows with the same timestamp into one, effectively reducing the number of records, outputting the same table-like object with an additional \textit{counter} column with the number of represented rows.

The data exploration in \autoref{ssub:multiread-issue} showed that the datasets contain multiple records for the same sensor within the same second. The aggregation step has a two-fold purpose: enabling a time-based learning and making the detection system more robust to measurement flooding. The learning based on past states is supported in later steps by selecting a fixed number of records for the machine learning. In this way, the machine can take into account the previous states of the system. We explore the performance of two approaches: event-based (not aggregated where all sensor readings are provided sequentially), and time-based (time-aggregated where the records represent a fixed time frame).

The aggregation is performed as illustrated in \autoref{tab:aggregation}. All records with the same timestamp are replaced by a single record.
A \textit{counter} column is added to indicate how many measurements each output record represents.

This paper implemented the strategy of keeping the first reading of each sensor with the given timestamp. The \textit{counter} column has not been used in training or testing of the models. Both decisions are open for further research.

\newcolumntype{R}[2]{%
    >{\adjustbox{angle=#1,lap=\width-(#2)}\bgroup}%
    l%
    <{\egroup}%
}
\newcommand*\rot{\multicolumn{1}{R{45}{1em}}}

\begin{table*}[!t]
	\caption{Example of the aggregation step on a combined dataset: only rows with ordinal "0" are conserved}
	\label{tab:aggregation}
	\scriptsize
	\centering
\begin{tabular}{c c@{}c@{}c|c c|c c|c c|c c c c|c c|c c|c c c } \\
ord.& type & label &\multicolumn{1}{c}{date \& time}& \multicolumn{2}{c}{Fridge} & \multicolumn{2}{c}{Garage} & \multicolumn{2}{c}{GPS} & \multicolumn{4}{c}{Modbus} & \multicolumn{2}{c}{Motion} & \multicolumn{2}{c}{Thermostat} & \multicolumn{3}{c}{Weather} \\
\hline
0&injection & 1 & \multirow{3}{1.2cm}{2019-04-25 \\ 10:01:30} & 10.9 & True & 0 & 0 & - & - & 12503 & 61055 & 62763 & 5173 & - & - & 28.8 & 1 & 28.0 & -6.6 & 56.5 \\
1&injection & 1 & 											  & 12.6 & True & 0 & 0 & - & - & 1335 & 40858 & 30413 & 59303 & - & - & - & - & - & - & - \\
2&injection & 1 & 											  & 5.8 & False & 0 & 0 & - & - & - & - & - & - & - & - & - & - & - & - & - \\ \hline
0&injection & 1 & \multirow{3}{1.2cm}{2019-04-25 \\ 10:01:31} & - & - & - & - & 82.9 & 92.7 & - & - & - & - & 0 & 0 & 25 & 0 & 45.2 & 9.5 & 12.8 \\
1&injection & 1 & 											  & - & - & - & - & 1.3 & 11.1 & - & - & - & - & 0 & 0 & 26.9 & 1 & 33.0 & 0.4 & 98.6 \\
2&injection & 1 & 											  & - & - & - & - & - & - & - & - & - & - & - & - & 25.8 & 1 & 42.7 & -0.5 & 29.6 \\ \hline
\end{tabular}
\end{table*}

\subsubsection{Segmentation} 
\label{ssub:segmentation}
This operation accepts a series of records and outputs a series divided in blocks of the given size.

Our method intends to leverage on the evolution of the sensor readings in time and therefore, temporal continuity must be conserved as much as possible. Furthermore, we need to make sure that training and test datasets are absolutely disjoint. As the \toniot{} dataset does not provide separate training and test sets, we segment the full dataset into chunks of a parametrisable size (originally 1000\,records, TT dataset could not be divided in that way to contain all classes, so 500\,records parameter was chosen instead). We argue that the more subsequent readings are kept together, the better we can capture continuous evolution of the system and the patterns of an attack.

\subsubsection{Partitioning} 
\label{ssub:partitioning}
Taking the blocks of rows as input, this step outputs two datasets composed of records from these blocks.

Partitioning decides whether a segment of the aggregated dataset records will be part of the training or the testing dataset. A random seed is stored to be able to reproduce the uniform distribution given the percentage driving the partitioning. A verification is performed to ensure that each class is well represented in both partitions. The resulting datasets \texttt{TT500} and \texttt{PR1000} are presented in \autoref{tab:ttdatasets}.

\begin{table}[!t]
	\caption{Datasets generated for experiments}
	\label{tab:ttdatasets}
	\centering
	\begin{tabular}{ l p{5cm} }
\hline
\multicolumn{2}{l}{Derived from TT dataset} \\ \hline
TT500 non-aggregated 
& Closest to \textit{Combined IoT dataset}  \cite{alsaedi2020ton_iot} based on TT dataset without aggregation \\ 

TT500 aggregated  
& TT dataset with segmentation of 500, aggregated to have only 1 point per 1\,s \\
\\
\hline
\multicolumn{2}{l}{Derived from PR dataset} \\
\hline
PR1000 aggregated 
& PR dataset with segmentation of 1000, aggregated only 1 point per 1 s \\

		\end{tabular}
\end{table}

\subsubsection{Imputation} 
\label{ssub:missing_values_strategies}
The imputation step accepts a set of records and outputs the same records modified in a way that there are no missing values.

To construct a picture of the system's status from multiple sources of sensor readings, we face their misalignment. Sensors have different reporting frequencies and periods of activity. In order to keep a maximal extent of the detection, we are interested in all recordings having at least one sensor emitting. When a sensor is not emitting any data (while others are emitting), it appears as having "missing values".

The processing of the missing values is worth of a particular attention. Imputing them without careful analysis may worsen the overall result as indicated by our results of DataWig imputation, reported in \autoref{tab:results}.
IoT devices, due to their embedded character, are prone to being affected by network saturation. Therefore, missing values or even duplicated readings can provide valuable information for network status inference.

The strategies we implemented to deal with missing values are summarised in \autoref{tab:imputations}: \textit{constant imputation}, \textit{fill-forward}, \textit{median or mode value} (according to the data type), \textit{miss}, and "\textit{DataWig}" that uses the eponymous toolbox\cite{datawig}. It trains a deep learning model to choose the best fit for each column based on other columns, our strategy implements option \texttt{datawig.SimpleImputer.complete()}.

\textit{Constant imputation} strategy replaces missing fields with a pre-defined arbitrary value (e.g. $-100$ or $0$). \textit{Fill-forward} strategy propagates the last available value. In order to fill all values, backward fill was used as well. \textit{Median and mode} imputation is a common practice, too. 
An experimental approach, dubbed \texttt{miss}, has been introduced by ignoring all values completely and to perform machine learning only on the mask of missing values, that is a binary array indicating 1 where a value is missing and 0 where a value was found. Finally, \textit{DataWig}'s \texttt{SimpleImputer} that analysed data and filled the missing values in an autonomous manner, by applying deep learning.

As three channels are available as inputs for the chosen CNNs, mixed strategies were designed where different channels used different strategies.

The explored strategies are summarised in \autoref{tab:imputations}. The first five rows contain pure strategies where all three channels receive the same inputs, equivalent to a greyscale picture. The remaining rows describe composed strategies to explore the cases when channels' strategies differ.

\begin{table}[!t]
	\renewcommand{\arraystretch}{1.3}
	\caption{Strategies for data imputation}
	\label{tab:imputations}
	\centering
	\begin{tabular}{ r @{ } p{5.7cm}}
		\hline
		Imputation & Description
		\\ \hline 
    \texttt{-const} & Set missing to a constant value ($-100$)
		\\
    \texttt{+const} & Set missing to a constant value ($100$)
		\\
    \texttt{miss} & Set missing to 1 and all available readings to 0
		\\
    \texttt{fill }& Set to the last available reading (set to the next if the missing value is the first one)
		\\
    \texttt{datawig}& Using DataWig\cite{datawig} tool to impute missing values
		\\
		\\
		\texttt{fill2$|$miss1} & First two channels use \texttt{fill}, the third uses \texttt{miss} \\
		
		\texttt{miss2$|$fill1} & First two channels use \texttt{miss}, the third uses \texttt{fill} \\
		
		\texttt{-const2$|$miss1} & First two channels use \texttt{-const}, the third uses \texttt{miss} \\
		\hline
	\end{tabular}
\end{table}

\subsubsection{Tensor Construction} 
 \label{ssub:value_encoding}
\begin{figure}[!htbp]
  \includegraphics[width=.8\columnwidth]{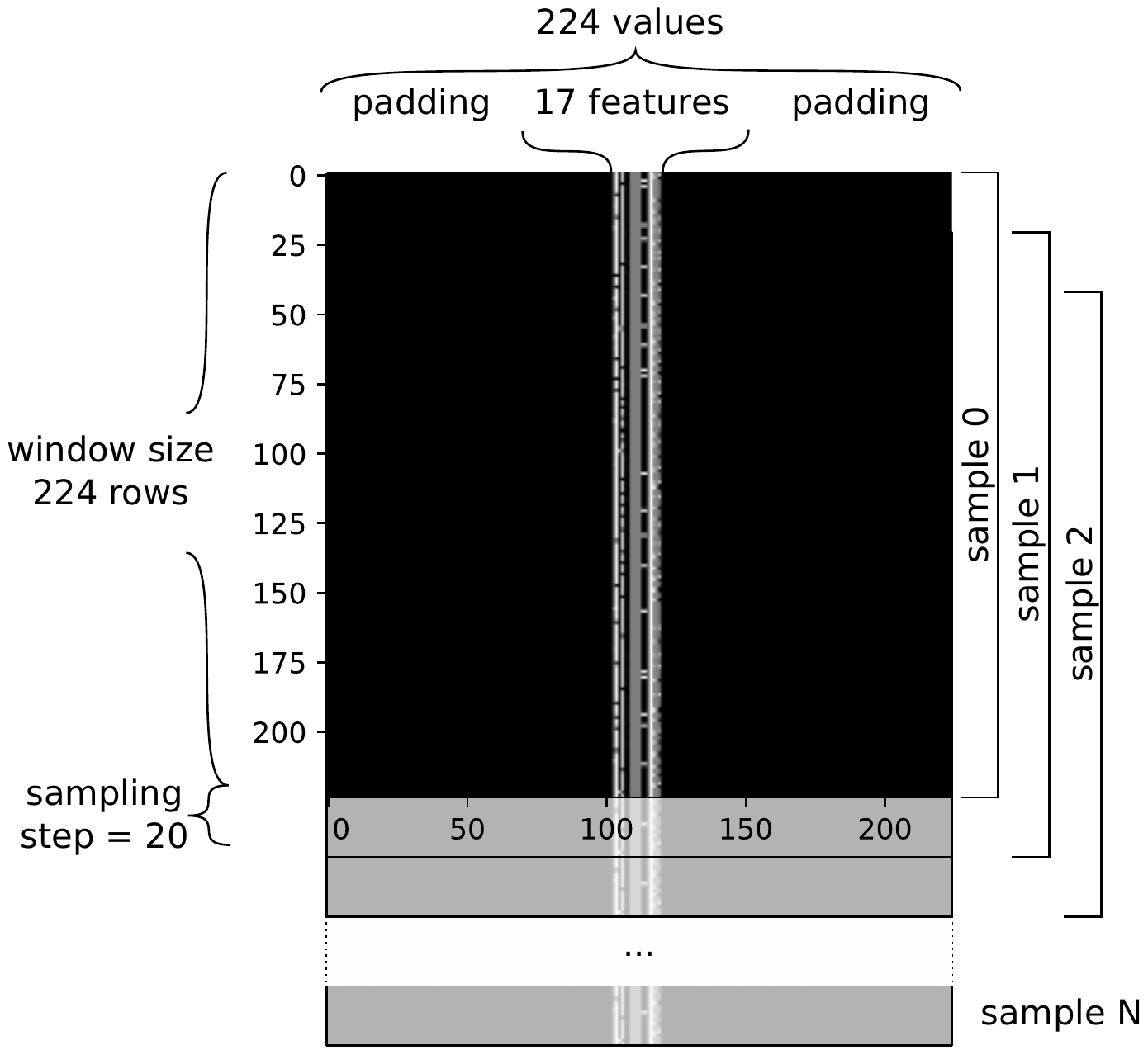}
  \centering
  \caption{Sampling of the data into a single-channel image. Example of \texttt{+const} strategy sample. As the actual width of data is 17 features, the most of the picture (left and right) is the padding.}
  \label{fig:sampling}
\end{figure}

The transformation of multivariate time-series vectors into a tensor data was performed in a straightforward manner, where the columns in the tensor are the readings from different sensors and the rows represent points in time. As the target CNN architectures (i.e. ResNet-50 and EfficientNet-B0) take a tensor of $224 \times 224 \times 3\,\textrm{channels}$, the input tensors were padded with zeros to reach the required shape (column count). In addition, each channel of the tensor refers to a particular imputation strategy. Sampling is detailed in \autoref{fig:sampling}. The \textit{step} parameter is a positive integer. The sample contains \num{224} records from our input. The following sample contains the same data but shifted by \textit{step} rows, adding \textit{step} more recent rows, thus forming a sliding window on our data. Lower values of \textit{step} produce more data samples with lower variation. However, based on our early experiments, the value of 20 reduces the training time without negatively affecting the outcome.

As a sample contains several annotated records, we need to determine the label for the sample. We adopted the strategy of the latest label appearing in the set of records used for the sample. The rationale is that the previous records are the past readings and only the last one is the current one, thus we chose its label to decide the ground truth.

This approach is applicable on TT and PR datasets as the labels are grouped in time and do not interfere (with the exception of scanning attack that is interlaced with password attack for the Motion Light sensor).

\section{Experiment design} 
\label{sec:experiment_design}
In this section, we present the experiments to evaluate the performance of our method. Specifically, the proposed data encoding method will be used to encode the inputs to two classical CNN models, i.e. ResNet \cite{he2016resnet} and EfficientNet \cite{tan2019efficientnet}, respectively. 
We will present the data preparation for the experiments and how to adapt the CNN models into our 8-class classification tasks. The unoptimised code was implemented by 
Python~3, and the experiments are running on a high-performance PC with CPU (
3.80\,GHz
) and Nvidia GeForce RTX 3070 GPU.

\subsection{Training and testing sets} 
\label{sub:training_and_testing_sets}

The first part of the pipeline, common to both CNNs, includes loading the data into the data structures, performing pre-processing steps according to \autoref{sub:data_processing}: combination, aggregation, separation in chunks of 500 or 1000 records, and their partitions as $70\%$ for training and $30\%$ for testing the models' performance. Resulting training and test datasets were transformed into an image-like formats according to mechanisms described in \autoref{ssub:value_encoding}.

The training and testing was performed on datasets based on \cite{alsaedi2020ton_iot} \textit{TT} dataset and \textit{PR} dataset. The \textit{TT} dataset becomes too small for an efficient machine learning (with only 864 samples of normal class with step=20, and 70 \% training set, with total of only \num{3148} samples within the training dataset) when aggregated. \textit{PR} dataset is the superset of \textit{TT} and contains more samples with the same set of features. The aggregated \textit{PR} dataset yields a similar amount of samples as the non-aggregated version of \textit{TT} as can be seen in \autoref{tab:results}.  
%

\subsection{CNNs} 
\label{sub:cnns}

The evaluated models are ResNet-50\cite{keras_ResNet50} and EfficientNet-B0\cite{keras_EfficientNetB0}, for which we use the pre-trained models
that %
has been built-in by 
Keras\cite{chollet2015keras}. Both architectures expect an input tensor of shape $224 \times 224 \times 3$. It is %
worthing to remark 
the EfficientNet architecture requires the inputs to be values between 0 and 255. On the other hand, during preliminary tests, ResNet showed equal perfomance when input value range is $[0, 1]$ and when it is $[0, 255]$. The reported experiments were conducted on the data scaled to the range of [0; 255].

The predictions on the test set were recorded in two ways: in an 8-class classification setting and derived binary classification. The derived binary classification is the one-against-all interpretation of 8-class classification: if any of attack classes is predicted, it is counted as a attack, otherwise a normal class is predicted.

The learning rates for both ResNet and EfficientNet are set as 
$10^{-2}$, and the training of the two models are 
running for $20$ epochs for relatively stable results, according to our early experiments.

\subsection{Metrics} 
\label{sub:metrics}

Results are assessed by True Positive Rate (TPR) and False Positive Rate (FPR) with regard to attack detection against the normal class.


In \cite{alsaedi2020ton_iot}, the authors use accuracy, precision, recall (also called sensitivity, or TPR). As our metrics focus on the probability of false alarm (FPR), a conversion of the metrics is required. The baseline FPR, best result of the binary classification task on combined dataset in \cite{alsaedi2020ton_iot} reported in \autoref{tab:results} was computed using the formula (\ref{eq:fpr-apt}). Its inference from accuracy, precision, recall, and FPR is beyond the scope of this paper.

\newcommand{\tpr}{\mathit{TPR}}
\newcommand{\fpr}{\mathit{FPR}}
\newcommand{\accuracy}{\mathit{accuracy}}
\newcommand{\precision}{\mathit{precision}}
\renewcommand{\tpr}{t}
\renewcommand{\fpr}{\mathit{FPR}}
\renewcommand{\accuracy}{a}
\renewcommand{\precision}{p}

\begin{equation}
\fpr = \frac{(1-\accuracy) (1- \precision) \tpr}{\precision \times (\accuracy - 2 \times \tpr) +  \tpr}
\quad
\text{where}
\begin{cases}
\accuracy &= \text{Accuracy} \\
\tpr &= \text{Recall (TPR)} \\
\precision &= \text{Precision} \\
\end{cases}
\\
\label{eq:fpr-apt}
\end{equation}





\section{Results} 
\label{sec:results}
In this section we have a closer look at the hyperparameters and we present the results of the training in form of charts and comparisons to the baseline in the original paper \cite{alsaedi2020ton_iot}.

\subsection{Explored Parameters}
Experiments were performed with several parameters: dataset, aggregation, imputation method, type of CNN. Their summary is in \autoref{tab:results}, our best results are highlighted in bold font.

\robustify\bfseries 
\robustify\itshape 

\begin{table}[!t]
	\caption{Results in different configurations at step=20, learning\_rate=0.01, epochs=20}
	\label{tab:results}
	\centering
	\begin{tabular}{
		l @{ }
		S[table-format=2.2, detect-weight, detect-shape, detect-mode] @{ }
		S[table-format=2.2, detect-weight, detect-shape, detect-mode] @{ }
		S[table-format=2.2, detect-weight, detect-shape, detect-mode] @{ }
		S[table-format=2.2, detect-weight, detect-shape, detect-mode] @{ }
		l @{ }
		r @{}
		S[table-format=5.0, detect-weight, detect-shape, detect-mode] @{}
	}
	\hline
& \multicolumn{2}{c}{ResNet-50} & \multicolumn{2}{c}{EfficientNet-B0}\\
Imputation & {TPR(\%)} & {FPR(\%)} & {TPR(\%)} & {FPR(\%)} & {Dataset} & {Samples} \\
\hline
-const3 & 83.15 & 2.42 & 90.04 & 3.30 & PR1000a & 14496 \\
-const3 & 92.14 & 21.52 & 92.93 & 52.02 & TT500a & 3148 \\
-const3 & 94.76 & 3.07 & 93.57 & 3.01 & TT500n & 14995 \\
\hline
-const2$|$miss1 & 89.36 & 3.07 & 90.11 & 3.37 & PR1000a & 14496 \\
-const2$|$miss1 & 96.63 & 25.56 & 87.54 & 20.18 & TT500a & 3148 \\
-const2$|$miss1 & 98.00 & 3.61 & 93.19 & 3.73 & TT500n & 14995 \\
\hline
+const3 & 79.55 & 3.40 & 91.69 & 4.32 & PR1000a & 14496 \\
\textbf{+const3} & 94.92 & 2.60 & \bfseries 94.86 & \bfseries 0.66 & \textbf{TT500n} & \bfseries 14995 \\
\hline
fill3 & 62.32 & 9.22 & 42.40 & 7.32 & PR1000a & 14496 \\
fill3 & 91.25 & 21.08 & 92.14 & 44.84 & TT500a & 3148 \\
fill3 & 85.68 & 2.70 & 78.00 & 7.15 & TT500n & 14995 \\
\hline
fill2$|$miss1 & 67.57 & 4.80 & 68.99 & 4.47 & PR1000a & 14496 \\
fill2$|$miss1 & 93.38 & 23.77 & 95.29 & 27.80 & TT500a & 3148 \\
fill2$|$miss1 & 93.46 & 14.68 & 94.65 & 15.96 & TT500n & 14995 \\
\hline
miss2$|$fill1 & 78.73 & 3.95 & 87.34 & 2.42 & PR1000a & 14496 \\
miss2$|$fill1 & 95.85 & 26.01 & 93.71 & 46.64 & TT500a & 3148 \\
miss2$|$fill1 & 98.65 & 10.76 & 96.22 & 7.18 & TT500n & 14995 \\
\hline
miss3 & 88.16 & 2.67 & 89.14 & 3.12 & PR1000a & 14496 \\
miss3 & 86.31 & 19.28 & 94.61 & 27.80 & TT500a & 3148 \\
\textbf{miss3} & \bfseries 92.43 & \bfseries 0.31 & 93.19 & 2.82 & \textbf{TT500n} & \bfseries 14995 \\
\hline
DataWig\cite{datawig} & 54.38 & 12.89 & 66.37 & 24.13 & PR1000a & 14496 \\
MM$^\mathrm{a}$ & 96.39 & 10.61 & 67.74 & 3.03 & TT500a & 3223 \\
\hline
\multicolumn{3}{l}{\textit{\cite{alsaedi2020ton_iot} CART$^{~\mathrm{b}}$:}}& \itshape 88 & \itshape 12\pm2$^{\mathrm{c}}$ & \textit{TT} & {-} \\
\multicolumn{3}{l}{\textit{\cite{alsaedi2020ton_iot} LSTM:}}& \itshape 81 & \itshape 19\pm2$^{\mathrm{d}}$ & \textit{TT} & {-} \\
\hline

\multicolumn{7}{p{.9\columnwidth}}{\rule{0pt}{1.5em}$^{\mathrm{a}}$ MM median and mode used in \cite{alsaedi2020ton_iot}}\\
\multicolumn{7}{p{.9\columnwidth}}{$^{\mathrm{b}}$ CART -- Classification and Regression Trees}\\
\multicolumn{7}{p{.9\columnwidth}}{$^{\mathrm{c, d}}$FPR computed using \autoref{eq:fpr-apt} on accuracy, precision, and recall, with uncertainty of 0.005 due to two-decimal-rounded data from \cite{alsaedi2020ton_iot}} \\
\multicolumn{7}{p{.9\columnwidth}}{$^{\mathrm{c}}$ for \textit{CART}, the $\fpr = 0.12\pm0.02$}\\
\multicolumn{7}{p{.9\columnwidth}}{$^{\mathrm{d}}$ for \textit{LSTM}, the $\fpr = 0.19\pm0.02$}
	\end{tabular}
\end{table}

Aggregation parameter indicates whether any merging of records was performed on the date and time attributes. We represent one moment in time by a fixed amount of data, i.e. one data point. In addition, to compare our results to the baseline presented in \cite{alsaedi2020ton_iot}, we implemented their approach without aggregation.

Imputation method represents the way of filling missing parts of records. Values were filled with a constant value (e.g. a value chosen out of the bounds, $-100$) and the special case of \texttt{miss} strategy that disregards values of the sensors and fills the features with \texttt{1} when the feature is missing, otherwise with \texttt{0}. This method generally outperforms the other imputation methods. The strategy DataWig is one of the worse performing. By imputing all missing values, the DataWig tool suppresses the information about what values were missing as its aim is to make the data as seamless as possible. Therefore, the poor performance of this strategy could indicate the importance of the pattern of missing values. \autoref{tab:imputations} defines the different tested strategies of data imputation.

The main feature of the ResNet is the ability to build up a very deep neural networks, where more than 100 layers can be stacked. The shortcut structure in ResNet diminishes the probability of gradient vanishing that was usually happening as the number of layers increases (i.e., the deeper networks) \cite{he2016resnet}. In these experiments, we use ResNet-50 that contains 50 convolutional layers for the high
efficient training and relatively satisfied 
performance. Similarly, EfficientNet family of CNNs is characterised by the effective scaling to tackle resource constraints\cite{tan2019efficientnet}. The EfficientNet-B0 is the smallest in number of parameters amongst the 8 (B0 to B7). It shows as the least performing evaluated on ImageNet\cite{deng2009imagenet}. As reported in \cite{tan2019efficientnet}, compared to ResNet-50, Efficient achieves top-1 accuracy at a similar level with 5 times less parameters, making it faster in training and testing.

\begin{figure}[!htbp]
  \includegraphics[width=.85\columnwidth]{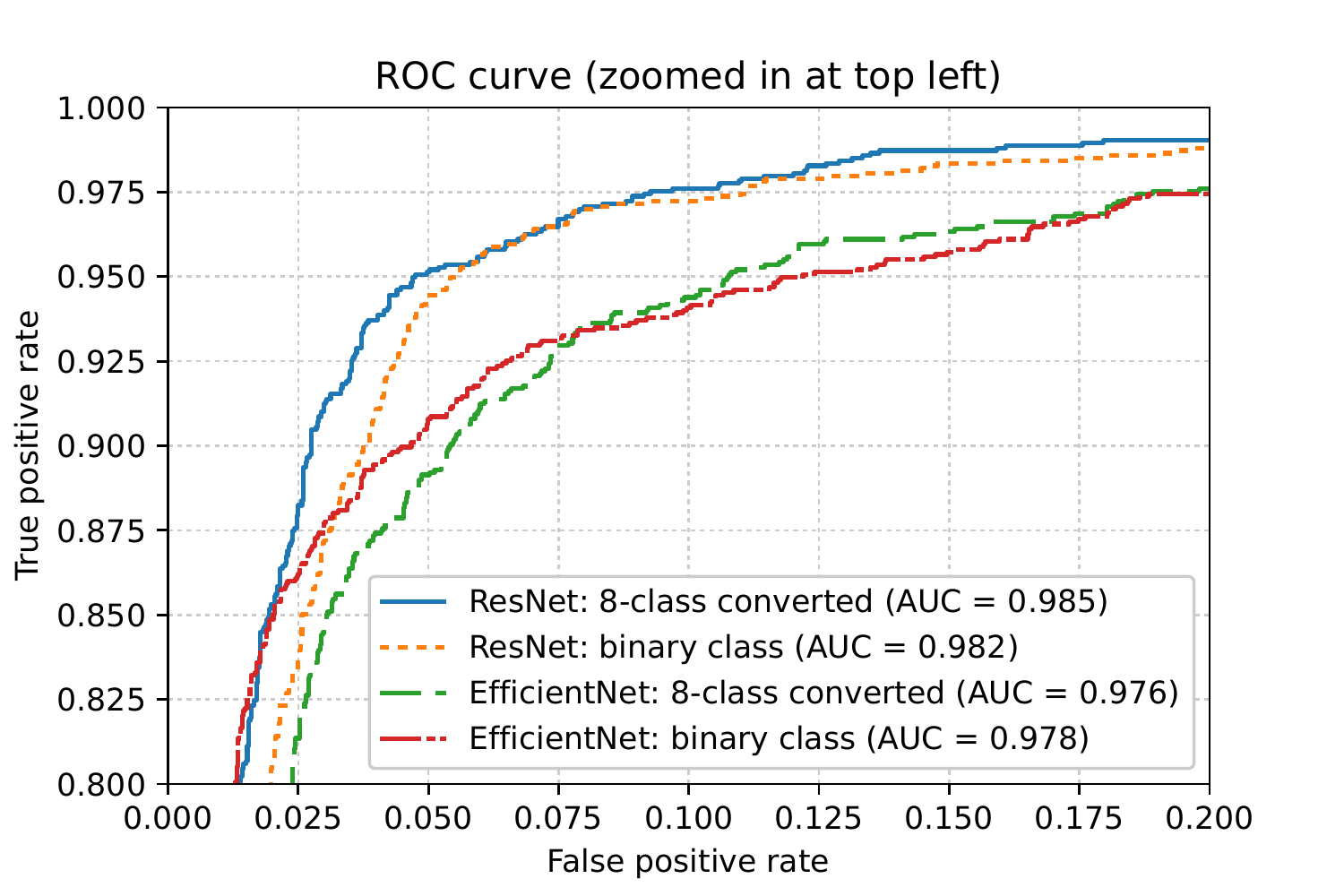}
  \centering
  \caption{Detail of Receiver Operating Characteristic curve and AUC values for EfficientNet (E) and ResNet (R) trained both as 8-class and binary classifiers on \texttt{PR1000} aggregated dataset}
  \label{fig:roc-binary-r-e-8-2-zoom}
\end{figure}

A detail of Receiver Operator Characteristic (ROC) curve of the imputation strategy \texttt{miss} experiment is represented in \autoref{fig:roc-binary-r-e-8-2-zoom} for ResNet and EfficientNet models. It was chosen for the illustration of relative performances of the binary classifier and the multi-class classifier. The models were trained as a binary classifier and as an 8-class classifier on \texttt{PR1000} non-aggregated dataset. The 8-class classifiers' curves represent the 1-class-against-all for normal class against all attack classes, the same way it was used to assess CNNs on the other imputation methods and datasets. The reported AUC values show that ResNet's 8-class classifier (AUC$=0.985$) out-performs its binary classifier (AUC$=0.982$) in this task. On the other hand, for EfficientNet, the binary classifier obtains slightly better AUC value of $0.978$ compared to 8-class classifier's $0.976$.

\subsection{Comparison of instances}

The results on the multi-class version of the evaluation are presented in the form of TPR, FPR measures including the training sample size (\autoref{tab:results}), in a setting of 8-class classifier converted to a binary classifier.


We observe that the best results were achieved on the non-aggregated TT dataset, i.e. \texttt{TT500} non-aggregated. Caution is to be exercised as the \texttt{normal} class is easier to detect in non-aggregated datasets. We attribute it to parts of the data with high multiplicity for one data point (observable in \autoref{fig:counts}). In this way, the tensor does not contain any missing values and therefore is easily identified as \texttt{normal}. 

Consistent low false positive rate is observed in \texttt{PR1000} aggregated dataset. It contains more data but without the redundant timestamps.

Poor results of \texttt{TT500} aggregated TT can be attributed to a small sample cardinality and the deep learning requiring more data to capture meaningful patterns. The least satisfying results were obtained on \texttt{PR1000} aggregated dataset using the \texttt{datawig} imputation strategy. 

Generally, ResNet-50 tends to perform better than EfficientNet-B0, with lower FPR. However, on the \texttt{TT500} non-aggregated dataset and the \texttt{+const} imputation strategy, the EfficientNet obtains one of the best results overall, whereas the best result for ResNet is with \texttt{TT500} non-aggregated dataset and \texttt{miss} imputation. Both results outperform the baseline presented in \cite{alsaedi2020ton_iot} that was achieved for binary classification using Classification and Regression Trees that outperformed LSTM.
Our experiments show that our method using ResNet with \texttt{miss} strategy is able to improve the TPR by $5.0\%$ while reducing the FPR by $97\%$ with respect to the reference baseline (CART). 
In the case of EfficientNet \texttt{+const} imputing strategy, improvement of the TPR reached $7.8\%$ and FPR decreased by $95\%$ compared to the best results of \cite{alsaedi2020ton_iot}. 

\begin{table}
	\caption{Confusion matrix of the best performing ResNet 8-class classifier on \texttt{TT500} non-aggregated dataset}
	\label{tab:confusion-matrixR}
	\centering
	\begin{tabular}{r @{ } c|r r r r r r r r}
		\multicolumn{2}{c|}{True Label}
					&	\multicolumn{8}{c}{Predicted Label} \\
					&   & b & d & i & n & p & r & s & x \\
		\hline
		backdoor	& b & 213 & 3 & 0 & 106 & 1 & 1 & 0 & 1 \\
		ddos		& d & 18 & 273 & 4 & 0 & 0 & 20 & 0 & 0 \\
		injection	& i & 0 & 0 & 521 & 9 & 2 & 1 & 2 & 0 \\
		normal		& n & 8 & 0 & 1 & 3179 & 0 & 0 & 0 & 1 \\
		password	& p & 0 & 8 & 7 & 19 & 364 & 1 & 5 & 4 \\
		ransomware	& r & 0 & 0 & 22 & 0 & 7 & 141 & 0 & 5 \\
		scanning	& s & 3 & 0 & 2 & 2 & 15 & 0 & 20 & 0 \\
		xss			& x & 0 & 0 & 0 & 4 & 1 & 0 & 0 & 45 \\
	\end{tabular}
\end{table}

\begin{table}
	\caption{Confusion matrix of the best performing EfficientNet 8-class classifier on \texttt{TT500} non-aggregated dataset}
	\label{tab:confusion-matrixE}
	\centering
	\begin{tabular}{r @{ } c|r r r r r r r r}
		\multicolumn{2}{c|}{True Label}
					&	\multicolumn{8}{c}{Predicted Label} \\
					&   & b & d & i & n & p & r & s & x \\
		\hline
		backdoor	& b & 271 & 1 & 0 & 47 & 3 & 3 & 0 & 0 \\
		ddos		& d & 19 & 237 & 3 & 7 & 0 & 49 & 0 & 0 \\
		injection 	& i & 0 & 0 & 529 & 5 & 0 & 0 & 1 & 0 \\
		normal 		& n & 3 & 0 & 3 & 3168 & 15 & 0 & 0 & 0 \\
		password 	& p & 0 & 1 & 5 & 33 & 356 & 13 & 0 & 0 \\
		ransomware	& r & 0 & 0 & 0 & 0 & 0 & 170 & 0 & 5 \\
		scanning	& s & 2 & 0 & 0 & 3 & 15 & 0 & 22 & 0 \\
		xss			& x & 0 & 0 & 0 & 0 & 3 & 0 & 0 & 47 \\
	\end{tabular}
\end{table}

The best performing models provided the confusion matrices in \autoref{tab:confusion-matrixR} and \autoref{tab:confusion-matrixE}. In the case of EfficientNet, we observe most classification errors as ddos was classified as ransomware and followed by backdoor misclassified as normal. For the ResNet instance, most errors concern backdoor misclassified as normal, while other errors are less significant.

\section{Discussion} 
\label{sec:discussion}
Two particular cases of wrong detections could be noticed: attack switched to a different type within the current window, a single-point different attack amidst of a uniform type.

In order to explain them, we look at the dataset and the sampling.
The implementation of the sliding window does not skip discontinuities. When the attack type changes, the most of the data indicates previous attack and it is understandable that the previous type is detected.

Similarly, a single-point attack detection is an outlier within a continuous block of the same-type attack. It is challenging or impossible to detect these with convolution networks, as a single row does not change the overall appearance of the image. A change in temporal representation could address this issue. For instance, more recent rows could be given more space, older rows less space. Newer ones could be duplicated to be noticed and older rows aggregated to provide a more stable history background.

Although we were able to obtain an important improvement of TPR and FPR (from $88\,\%$ to $92.43\,\%$ and from $12\,\%$ to $0.66\,\%$ respectively) compared to the baseline from \cite{alsaedi2020ton_iot}, we would like to point out limitations of this method.

\toniot{} datasets lack repetitions of the same attack, the observations contain significant time gaps, the time without observations is not labelled.

The patterns that our method captures predict different attacks. However, they might be an indication of the current configuration or performance of the sensor simulation. \autoref{fig:composition-tt} shows that only scanning attack appears in the dataset in two instances -- other attacks appear only in uninterrupted blocks. For example, all events and only events on 27-Apr-19 between 04:57:36 and 05:40:53 are labelled as XSS. 

In addition, no transition from normal state to any instance of the attack appears in the data. Normal observations are separated from attack scenarios by an observation gap more than 20 days long.

As the recordings of the data during attacks are concentrated within a relatively short period of time (hours), we have no data to confirm that the same attack on the next day could be detected.
In other words, as there is little variability, the model might as well learn ephemeral patterns that depend more on the time of day than on what actually happens in the network.

Therefore, we cannot rule out that the system performs well on this dataset only because the dataset is flawed in a way that produces specific patterns for each segment of time. Multiple scenarios of the same type of attack could immensely improve the confidence in the tested methods.

A reliable comparison to the baseline is challenging as the \toniot{} dataset does not provide separate training and test sets.

Furthermore, the paper introducing the \toniot{} dataset\cite{alsaedi2020ton_iot} does not specify whether the attacks have an actual influence on either values or on any meta-information that is available via this dataset.

A future extension of our experimentation could explore simulated removal, addition or replacement of sensors. The approach is potentially scalable to 224 features as only 17 slots out of 224 were used in the current experimentation, for higher levels of EfficientNet and ResNet, more features can be used.


These observations may contribute to future dataset evaluation methods. Amidst calls for more real-time IoT datasets \cite{malhotra2021iot-challenges}, a framework for their evaluation could benefit potential data providers during their testbed design and data collection considerations.

\section{Conclusion}
In the context of cybersecurity challenges in IoT and IIoT, this paper explores novel possibilities of deep learning applications on sensor readings to determine the intrusion state of the system.
We deployed two CNN architectures on the same data to observe how their performance differs. The deployed architectures are ResNet-50 and EfficientNet-B0.
We presented an approach to represent sensor readings for use with CNNs that takes into account past sensor readings as the context for the current prediction. The time axis corresponds to the vertical axis in the encoding of the image to be processed by CNNs.
We also showed how the presence or absence of values in the dataset can influence the intrusion detection of the monitored system. Overall results in binary classification show a significant improvement compared to the baseline\cite{alsaedi2020ton_iot}. Leveraging on patterns of missing values in time, false alarm rate fell from 12\,\% in the best baseline to 0.3\,\% or 0.7\,\% with ResNet or EfficientNet respectively, while the recall (true positive rate) raised from 88\,\% to 92\,\% or 94\,\% respectively.
\bibliographystyle{sty/IEEEtran}
\bibliography{main}

\end{document}